\documentclass[aps,prl,twocolumn,showpacs,showkeys,groupedaddress,amsmath,twoside,footinbib]{revtex4-1}

\usepackage{graphicx,dsfont,natbib,subfigure}

\newcommand{\Xp}{X^{\mathrm{\scriptscriptstyle{(+)}}}}

\newcommand{\vev}[1]{\langle #1 \rangle}

\newcommand{\copN }[2][]{\mathcal{C}_{#1}#2} 
\newcommand{\CopN }[2][]{\mathcal{C}_{#1}\!\left(#2\right)} 
  
\newcommand{\ribracket}{] \mspace{-3 mu} ]}
\newcommand{\libracket}{[\mspace{-3 mu} [}

\begin{document}


\title{A model-free characterization of recurrences in stationary time series}


\author{R\'emy~Chicheportiche}
\email{remy.chicheportiche@ecp.fr}
\author{Anirban Chakraborti}
\email{anirban.chakraborti@ecp.fr}
\affiliation{
Chaire de finance quantitative, \'Ecole Centrale Paris, 92\,295 Ch\^atenay-Malabry, France
}


\date{\today}

\begin{abstract}
Study of recurrences in earthquakes, climate, financial time-series, etc.\ is crucial to better forecast disasters and limit their consequences. 
However, almost all the previous phenomenological studies involved only a long-ranged autocorrelation function, 
or disregarded the multi-scaling properties induced by potential higher order dependencies. 
Consequently, they missed the facts that
  non-linear dependences \emph{do} impact both the statistics and dynamics of recurrence times, and
 that scaling arguments for the unconditional distribution may not be applicable.
We argue that copulas is the correct model-free framework to study non-linear dependencies in time series and related concepts like recurrences. 
Fitting and/or simulating the intertemporal distribution of recurrence intervals is very much system specific, 
and cannot actually benefit from universal features, in contrast to the previous claims. 
This has important implications in epilepsy prognosis and financial risk management applications.
\end{abstract}

\keywords{Recurrence intervals, copulas, long-ranged correlations, time series}

\maketitle


Extreme events are widely studied in seismicity, astronomy, physiology, finance, etc. \cite{bunde2002science}
In particuliar, the inter-occurence times (or recurrence intervals), 
i.e.\ the periods between two occurences of the observed phenomenon that exceed a given threshold, 
have important implications in risk management in a view to
predict the advent of such extreme events or characterize aftershocks.

A consensus has emerged on the statistics of recurrence times
from many phenomenological studies 
on real or simulated data with long-ranged correlations: 
the unconditional probability distribution function (PDF) of recurrence intervals $\tau$ 
follows a scaling relation \cite{bak2002unified,bunde2005long,*PhysRevE.75.011128,PhysRevE.78.051113} 
\begin{equation}\label{eq:scaling}
    \pi(\tau)= \frac{1}{\tau_{\text{c}}}\,f\!\left(\frac{\tau}{\tau_{\text{c}}}\right),
\end{equation}
where $\tau_{\text{c}}$ is a characteristic recurrence time for a given time series and a choice of event-triggering threshold.
Possible and reported scaling functions include:
exponential decay $\ln f(x)\sim x$  (which corresponds to independent arrivals),
power-law decay $\ln f(x)\sim \ln x$, 
stretched exponential (generalized Gamma) $\ln\ln f(x)\sim \ln x$, 
and others, e.g.\ mixed stretched exponential + power-law, 
possibly due to finite size and discreteness effects \cite{PhysRevE.75.011128,PhysRevE.78.051113}.
Furthermore, the sequence of recurrence times $\{\tau_i\}$ exhibits long-term correlation $\vev{\tau_i\tau_{i+\ell}}\sim\ell^{-\gamma}$ \cite{PhysRevLett.95.208501}.
Yet we show that this consensus, founded on the misconception that 
``nonlinearities are not needed to explain the properties studied'' \cite{bunde2005long},
is incorrrect since the whole non-linear dependences affect $\pi(\tau)$. 
In particular, simple theories based on a benchmark model of triggered seismicity have also found this universality 
to be only approximate \cite{saichev2006universal,*touati2009origin}.

Some {\it ad hoc} attempts at bringing in non-linearities and/or multi-scaling have also been made, 
in view of modelling the behavior of a specific system, e.g.\ financial returns with multifractal log-volatility 
\cite{PhysRevLett.99.240601}. 
In this paper we show that copulas is the correct \emph{model-free} theoretical framework to study non-linear dependencies in time series.
This implies that non-linear correlations and multi-point dependences are relevant for the related concept of recurrences.
As a consequence, a scaling relation of the form \eqref{eq:scaling} is at best {approximate}, 
and would only hold for processes exhibiting a time-dependence characterized by a unique time scale $\tau_{\text{c}}$. 
Furthermore, a characterization of clustering based on the autocorrelation of recurrence intervals is an oversimplified view of the reality.

\paragraph{}
We consider a time series $\{X_t\}_{t=1\ldots T}$ of length $T$, 
 as a realization of a discrete stochastic process.
The joint cumulative distribution function (CDF) of $n$ occurrences ($1\leq t_1<\ldots<t_n<T$) of the process is 
\begin{equation}
    \mathcal{F}_{t_1,\ldots,t_n}(\mathbf{x})=\mathds{P}[X_{t_1}<x_{t_1},\ldots,X_{t_n}<x_{t_n}].
\end{equation}
We assume that the process is stationary with a distribution $F$,
and a translational-invariant joint distribution $\mathcal{F}$ with long-ranged dependences, 
as is typically the case e.g.\ for seismic and financial data.

A realization of $X_t$ at date $t$ will be called an ``event'' when its value exceeds a threshold \mbox{$\Xp=F^{-1}(1\!-\!p_+)$}
that corresponds to the upper \mbox{$p_+$-quantile} of the marginal distribution.
As we show just below, 
recurrences of such events only involve the diagonal $n$-points probability 
\begin{equation}\label{eq:CopN}
    \CopN[n]{p}=\mathcal{F}_{t+\libracket 1,n\ribracket}(F^{-1}(p),\ldots,F^{-1}(p)),
\end{equation}
that all $n\geq 1$ consecutive variables $X_{t+1},\ldots,X_{t+n}$ are below 
the upper \mbox{$p$-th} quantile of the stationary distribution,
where $p\in[0,1]$ and $t+\libracket 1,n\ribracket$ is a shorthand for $\{t\!+\!1,\ldots,t\!+\!n\}$.
Clearly, $\CopN[1]{p}=p$ and we set by convention $\CopN[0]{p}\equiv 1$.
Eq.~\eqref{eq:CopN} in fact defines, through Sklar's theorem, the diagonal of what statisticians call the $n$-points ``copula'',
which is nothing else than a multivariate CDF with uniform marginals, see e.g.\ Ref.~\cite{ruschendorf2013copulas}.

As an example, the Gaussian diagonal copula is
\begin{equation}\label{eq:GaussCop}
    \CopN[n]{p}=\Phi_\rho\big(\Phi^{-1}(p),\ldots,\Phi^{-1}(p)\big)
\end{equation}
where $\Phi^{-1}$ is the univariate inverse CDF, and $\Phi_\rho$
denotes the multivariate CDF with $(n\times n)$ covariance matrix $\rho$,
which is T\oe{}plitz with symmetric entries
\begin{equation}\label{eq:2points}
    \rho_{tt'}\equiv\vev{X_tX_{t'}}=\rho(|t-t'|),\quad t,t'=1,\ldots,n.
\end{equation}
Although the $n$-points expectations of Gaussian processes reduce to all combinations of the $2$-points expectations~\eqref{eq:2points},
their full dependence structure is \emph{not} reducible to the bivariate distribution, 
unless the process is also  Markovian, i.e.\ only in the particular case of exponential correlation.
The White Noise product copula
 $\CopN[n]{p}=p^n$ is recovered in the limit of vanishing correlations $\rho(\ell)=0\, \forall\ell$, and
other examples include the exponentially correlated Markovian Gaussian Noise, 
the power-law correlated (thus scale-free) Fractional Gaussian Noise, and the logarithmically correlated multifractal Gaussian Noise.

Copulas are invariant under any continuous monotonous transformation of the $X_t$'s, 
and are thus better suited to study temporal dependences than e.g.\ the linear correlation function.
Indeed if $f$ and $g$ are two increasing functions, 
$\vev{f(X_t)f(X_{t+\tau})}$ and $\vev{g(X_t)g(X_{t+\tau})}$ can be arbitrarily different
in spite of the underlying process being the same,
whereas the copulas of $\{f(X_t)\}_t$ and $\{g(X_t)\}_t$ are identical.

Empirically, the $n$-points probabilities are very hard to measure due to the large noise associated
with such rare joint occurences.
However, there exist observables that embed many-points properties and are more easily
measured, such as the length of sequences (clusters) of thresholded events \cite{RC_AC_2013},
and the recurrence times of such events, that we study here.

\paragraph{}
The probability $\pi(\tau)$ of observing a recurrence interval $\tau$ between two events 
is the conditional probability of observing a sequence of $\tau-1$ ``non-events'' bordered by two events:
\[
    \pi(\tau)=\mathds{P}[X_{\tau}\!>\!\Xp, X_{\libracket 1;\tau\libracket}\!<\!\Xp|X_{0}\!>\!\Xp].
\]
After a simple operation flipping all `$>$' signs to `$<$', 
it can be written in the language of copulas as~\cite{chicheportiche2013phd}:
\begin{align}\label{eq:distrecint}
        \pi(\tau)=\frac{\CopN[\tau-1]{1\!-\! p_+}-2\,\CopN[\tau]{1\!-\! p_+}+\CopN[\tau+1]{1\!-\! p_+}}{p_+}.
\end{align}
The cumulative distribution
\[
    \Pi(\tau)=\sum_{n=1}^\tau \pi(n)=1-\frac{\CopN[\tau]{1\!-\! p_+}-\CopN[\tau+1]{1\!-\! p_+}}{p_+}
\]
is more appropriate for empirical purposes, being less sensitive to noise.%
\footnote{If $\pi(\tau)$ has exponential tails, then $1-\Pi(\tau)$ has the same tails;
          if $\pi(\tau)$ has power-law   tails, then $1-\Pi(\tau)$ has power-law tails, too.
          Hence representing $1-\Pi(\tau)$ in a lin-log or log-log scale, respectively, 
          is as meaningful as representing $\pi(\tau)$ itself.}
These exact expressions make clear --- almost straight from the definition --- that 
  (i)~the distribution of recurrence times \emph{depends only on the copula} of the underlying process 
      and not on the stationary law, in particular its domain or its tails 
      (this is because we take a relative definition of the threshold as a quantile); 
 (ii)~\emph{non-linear} dependences are highly relevant in the statistics of recurrences, so that 
linear correlations can in the general case by no means explain alone the properties of $\pi(\tau)$ \cite{altmann2005recurrence}; and 
(iii)~recurrence intervals have a \emph{long memory} revealed by the $(\tau\!+\!1)$-points copula being involved, so that 
only when the underlying process $X_t$ is Markovian will the recurrences themselves be memoryless.%
\footnote{It may be mentioned that in a non-stationary context, both renewal processes and the fractional Poisson process are also able to produce independent consecutive recurrences \cite{PhysRevE.78.051113,Sazuka20092839,politi2011full}.}
 Hence, when the copula is known (Eq.~\eqref{eq:GaussCop} for Gaussian processes), the distribution of recurrence times is exactly characterized by the analytical expression in Eq.~\eqref{eq:distrecint}.


The average recurrence time $\mu_\pi\equiv\vev{\tau}$ is found straightforwardly, 
and the variance $\sigma_\pi^2\equiv\vev{\tau^2}-\mu_\pi^2$ of the distribution can be computed as well:
\begin{align}
\label{eq:avrecint}
    \mu_\pi&=\frac{1}{p_+},\\ 
\label{eq:secmon}
    \sigma_\pi^2&=\frac{2}{p_+}\sum_{\tau=1}^{\infty}\CopN[\tau]{1\!-\!p_+}-\frac{1\!-\!p_+}{p_+^2}.
\end{align}
Importantly, $\mu_\pi$ is \emph{universal} whatever the dependence structure.%
\footnote{This result was first stated and proven by \citet{kac1947notion}, in a similar fashion.}
Introducing the copula allows to emphasize the validity of the statement even in the presence of non-linear long-term dependences,
as Eq.~\eqref{eq:avrecint} means that the average recurrence interval is \emph{copula}-independent.
This is intuitive as, for a given threshold, the whole time series is the succession of a fixed number $p_+T$ of recurrences 
whose lenghts $\tau_i$ necessarily add up to the total size $T$, so that $\vev{\tau}=\sum_i\tau_i/(p_+T)=1/p_+$.
Note that Eq.~\eqref{eq:avrecint} assumes an infinite range for the possible lags $\tau$, which is achieved either 
by having an infinitely long time series, or more practically when the translational-invariant copula is periodic at the boundaries of the time series, as
is typically the case for artificial data which are simulated using numerical Fourier Transform methods.
The variance $\sigma_\pi^2$ is not universal, in contrast with the mean, 
and can be related to the average unconditional waiting time \cite{RC_AC_2013}.

The universality of the average recurrence time has important implications for the potential scaling properties of the PDF of recurrence times.
Indeed, it has been believed that $\pi(\tau)$ can be fitted by a unique scaling function Eq.~\eqref{eq:scaling},
with $f$ depending on the underlying process and its dependence structure.
Such a scaling would be of paramount importance in the empirical investigation of extreme events
(for which $p_+$ is close to $0$ and there is often too few data points to conduct a thorough statistical study),
as it would make it possible to extrapolate the distribution found at low thresholds to large ones.
Now, because there is a one-to-one correspondance between $\vev{\tau}$ and $p_+$ following the universality \eqref{eq:avrecint}, 
the natural scale $\tau_{\text{c}}$  is necessarily the mean recurrence times, and
the relation \eqref{eq:scaling} describes in fact a scaling of $\pi$ with the threshold.

On the theoretical side, the fact that the distribution of recurrence times exhibits a universal scaling 
ranging over several orders of magnitude has been shown to be related to criticality in long-ranged correlated complex systems, 
like invariance under Renormalization-Group transformation \cite{PhysRevLett.95.028501,*corral2009point}.
But such a universal scaling can only exist for \emph{linearly} correlated processes \cite{altmann2005recurrence},
and there is even a concern \cite{molchan2005interevent} whether there exist at all fixed-point solutions of the RG equations 
other than the trivial exponential function corresponding to independent arrivals.
Therefore, since the average recurrence time does not carry any information whatsoever about the dependence structure of the process,
a scaling relation of the form \eqref{eq:scaling} is a trivial one when $\tau_{\text{c}}=\vev{\tau}$,
and it might hold for processes exhibiting a time-dependence fully characterized by linear correlations, 
and no other relevant time scale; 
this is the case only for the power-law correlated Gaussian process \cite{newell1962zero},
for which $\pi(\tau)$ can anyways be expressed analytically according to Eqs.~(\ref{eq:GaussCop},\ref{eq:distrecint}) with $\rho(\ell)\sim \ell^{-\gamma}$.
But no such simple scaling is expected in the general case when either non-linear dependences are present, 
and/or when several {time scales} are involved in the dependence structure.

\begin{figure}
    \center
    \subfigure[IBM daily stock returns.]
                                        {\label{fig:tailprobIBM}\includegraphics[scale=.8,trim=0 0 290 0,clip]{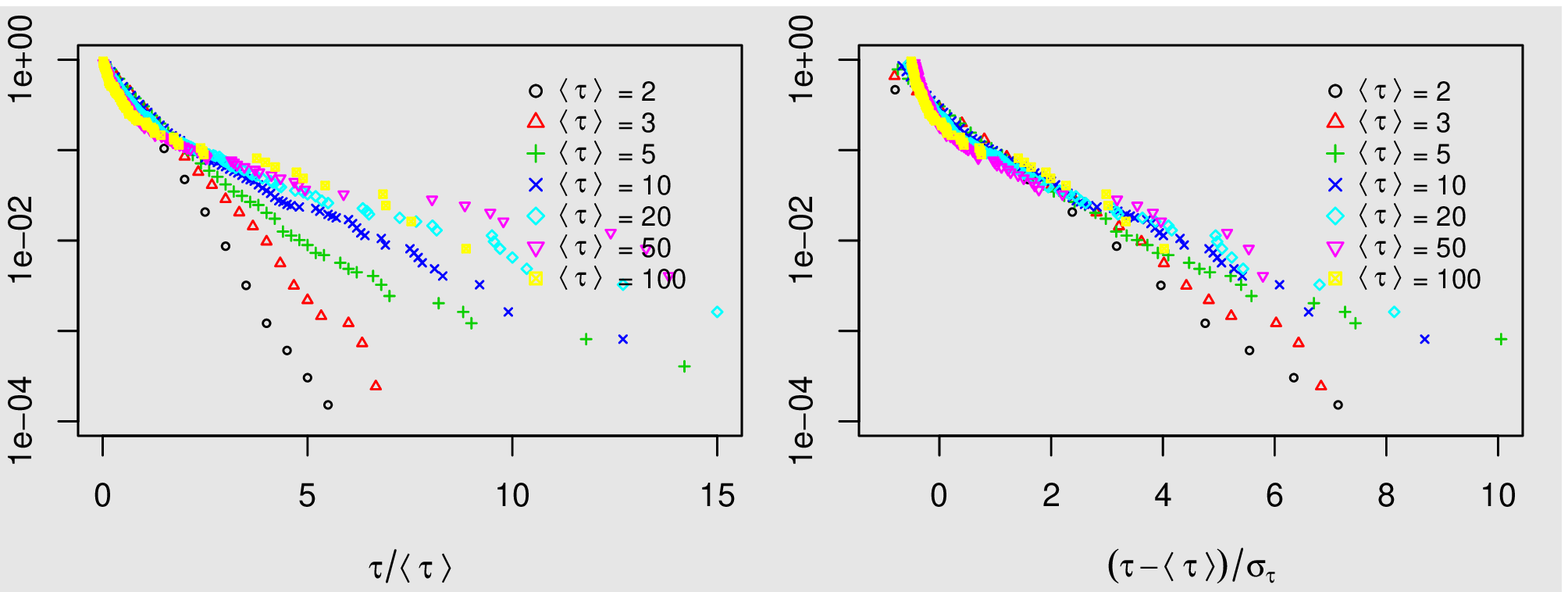}}
    \subfigure[EEG]{\label{fig:tailprobEEG}\includegraphics[scale=.8,trim=0 0 290 0,clip]{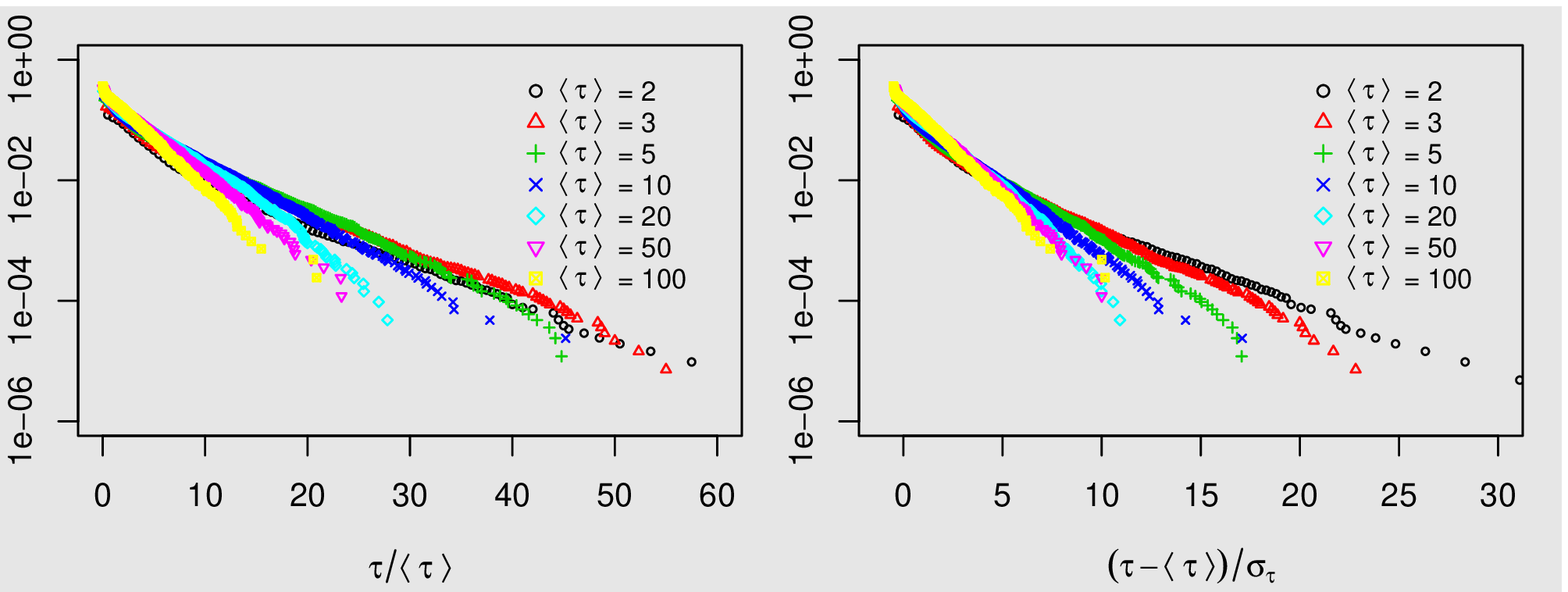}}
    \caption{\!(color online) To illustrate that no scaling occurs when non-linear dependences and/or several {time scales} are involved in the dependence structure, 
    we plot the tail probability $1-\Pi(\tau)$ of the recurrence intervals versus $\tau/\vev{\tau}$, 
    at several thresholds $p_+=1/\vev{\tau}$ for (a) financial data, and (b) EEG data.}
    \label{fig:tailprob}
\end{figure}

We illustrate this on Fig.~\ref{fig:tailprobIBM} for the daily log-returns of the IBM stock from 1962 to 2010 
(same data as in \cite{ludescher2011universal}, with however a sign flip since our definition of the threshold is upside whereas there it is downside).
Financial returns have a very short term linear correlation, 
a very long term and multifractal quadratic correlation, 
and other kinds of intermediate-scale dependences (e.g.\ leverage effect) \cite{perello2004multiple}, 
that show up in the copula in a very peculiar fashion \cite{chicheportiche2011goodness}.
Hence, as expected \cite{ludescher2011universal}, the distribution of recurrence times has a more complex 
functional form than what is allowed by the scaling in Eq.~\eqref{eq:scaling}, as reproduced on Fig.~\ref{fig:tailprobIBM}.
One may note by the way that the correlation of the amplitudes (volatility clustering) is the dominant one, 
so that the series of $\omega_t=\log|X_t|$ is mostly linearly correlated and
in fact the recurrence intervals distribution of $\omega_t$ does exhibit an approximate scaling of the type \eqref{eq:scaling} \cite{yamasaki2005scaling,*wang2006scaling,PhysRevE.75.011128}.

We also illustrate the non-scaling behavior on measurements of 
electroencephalograms (EEG) on brain surface of patients awake with eyes open (set A of the dataset studied in Ref.~\cite{PhysRevE.64.061907}).
It consists of 100 stationary series of 4097 observations each; 
once centered and rescaled, all series have same distribution.
The results are shown on Fig.~\ref{fig:tailprobEEG}: the presence of several (in fact periodic) scales in
the dependence structure forbids the scaling of $\pi(\tau)$ with $\vev{\tau}$.

In passing, notice that the statistics of recurrence times is much related to that of sequence lengths: 
an interval $\tau$ between two events always characterizes at the same time a sequence of $\tau-1$ ``non-events''.
In this respect, the average sequence length can also be shown to be universal, what rules out part of the analysis of Ref.~\cite{boguna2004conditional},
where the authors use the relation \eqref{eq:avrecint} as a \emph{test} of independence of the events above/below the threshold $\Xp=0$.

\paragraph{}
The dynamics of recurrence times is as important as their statistical properties,
and in fact impacts the empirical determination of the latter
\footnote{Distribution testing for $\pi(\tau)$ involving Goodness-of-fit tests \cite{ren2010recurrence} 
should be discarded because those are not designed for dependent samples and rejection of the null
 cannot be relied upon. See Ref.~\cite{chicheportiche2011goodness} for an extension of GoF tests when some dependence is present. 
}.
It is now clear, both from empirical evidences and analytically from the discussion on Eq.~\eqref{eq:distrecint}, 
that recurrence intervals have a long memory.
In dynamic terms, this means that their occurences show some clustering. The natural question is then:
``Conditionally on an observed recurrence time, what is the probability distribution of the next one?''
This probability  of observing an interval $\tau'$ immediately following an observed recurrence time $\tau$
is
\begin{widetext}
\begin{equation}
    \mathds{P}[X_{\tau+\tau'}>\Xp, X_{\tau+\libracket 1;\tau'\libracket}<\Xp|X_\tau>\Xp, X_{\libracket 1;\tau\libracket}<\Xp,X_0>\Xp].
\end{equation}
\end{widetext}
Again, flipping the `$>$' to '$<$' allows to decompose it as
\[
     \frac{\copN[\tau-1;\tau'-1]{}-\copN[\tau;\tau'-1]{}-\copN[\tau-1;\tau']{}+\copN[\tau;\tau']{}}{\copN[\tau-1]{}-2\copN[\tau]{}+\copN[\tau+1]{}}
    -\frac{\pi(\tau+\tau')}{\pi(\tau)},
\]
where the $(\tau\!+\!\tau')$-points probability
\[
   \CopN[\tau;\tau']{p}=\mathcal{F}_{\libracket 0;\tau\!+\!\tau'\ribracket\backslash\{\tau\}}(F^{-1}(p),\ldots,F^{-1}(p))
\]
shows up.
Of course, this exact expression has no practical use, 
again because there is no hope of empirically measuring any many-points 
probabilities of extreme events with a meaningfull signal-to-noise ratio.
We rather want to stress that non-linear correlations and multi-points dependences are relevant, 
and that a characterization of clustering based on the autocorrelation 
of recurrence intervals is an oversimplified view of reality.

\paragraph{}
The exact universality of the mean recurrence interval imposes a natural scale in the system.
A scaling relation in the distribution of such recurrences is only possible in the absence of any other characteristic time.
When such additional characteristic times are present (typically in the non-linear correlations), 
the rescaling must be performed with a scale-dependent renormalization, in the spirit of 
\begin{equation}\label{eq:scaling2}
    \pi(\tau)=\frac{1}{\sigma_\tau}g\!\left(\frac{\tau-\vev{\tau}}{\sigma_\tau}\right),
\end{equation}
where $\sigma_\tau$ is defined in Eq.~\eqref{eq:secmon} and \emph{does} contain information about
the dependence structure.
Then in some cases where $\sigma_\tau$ embeds all the dependence of the underlying process,
 the scaling \eqref{eq:scaling2} might hold, at least approximately, even in the presence of several scales, 
as the recurrence times corresponding to the different regimes would be renormalized accordingly.

We stress that recurrences are intrinsically multi-points objects 
related to the non-linear dependences in the underlying time-series.
As such, their autocorrelation is not a reliable measure of their dynamics,
for their conditional occurence probability is much history dependent.
We report elsewhere \cite{RC_AC_2013} more properties of recurrence times and the statistics of other 
observables (waiting times, cluster sizes, records, aftershocks) in light of their decription in terms of the diagonal copula.
We hope that these studies can shed light on the $n$-points properties of the process by 
assessing the statistics of simple variables
rather than positing an {\it a priori} $2$-points correlation structure and deriving a
 corresponding recurrence times distribution.

\acknowledgements
We thank F.~Abergel, F.~Baldovin, J.-P.~Bouchaud, D.~Challet, A.~Corral, N.~Millot, M.~Politi, E.~Scalas, G.~Tilak for helpful discussions and comments. 
R.~C. acknowledges financial support by Capital Fund Management, Paris.

\bibliography{../biblio_ECP}
\nocite{kac1947notion,Sazuka20092839,politi2011full}

\clearpage\appendix
\begin{widetext}

  \newcommand{\esp   }[1]{\mathds{E}[#1]}
  \newcommand{\pr    }[1]{\mathds{P}[#1]}
  \newcommand{\1     }[1]{\mathds{1}_{\{#1\}}}
  \newcommand{\Esp   }[1]{\mathds{E}\!\left[#1\right]}
\renewcommand{\Pr    }[1]{\mathds{P}\!\left[#1\right]}

\subsubsection*{Equation~(6)}
\[
    \pi(\tau)=\pr{X_{\tau}>\Xp, X_{\libracket 1;\tau\libracket}<\Xp|X_{0}>\Xp}.
\]
After a simple algebraic transformation flipping all `$>$' signs to `$<$', 
it can be written in the language of copulas as:
\begin{align}\nonumber
        \pi(\tau)&=\frac{\pr{X_{\tau}>\Xp, X_{\libracket 1;\tau\libracket}<\Xp; X_{0}>\Xp}}{\pr{X_{0}>\Xp}}\\\nonumber
                 &=\frac{\pr{X_{\libracket 1;\tau\libracket}<\Xp}}{p_+}
                  -\frac{\pr{X_{\tau}<\Xp, X_{\libracket 1;\tau\libracket}<\Xp}}{p_+}\\\nonumber
                 &-\frac{\pr{X_{\libracket 1;\tau\libracket}<\Xp; X_{0}<\Xp}}{p_+}
                  +\frac{\pr{X_{\tau}<\Xp, X_{\libracket 1;\tau\libracket}<\Xp; X_{0}<\Xp}}{p_+}\\\tag{6}
        \pi(\tau)&=\frac{\CopN[\tau-1]{1\!-\! p_+}-2\,\CopN[\tau]{1\!-\! p_+}+\CopN[\tau+1]{1\!-\! p_+}}{p_+}.
\end{align}

\subsubsection*{Equation~(9)}
The probability  of observing an interval $\tau'$ immediately following an observed recurrence time $\tau$
is
\begin{equation}\tag{9}
    \pr{X_{\tau+\tau'}>\Xp, X_{\tau+\libracket 1;\tau'\libracket}<\Xp|X_\tau>\Xp, X_{\libracket 1;\tau\libracket}<\Xp,X_0>\Xp}.
\end{equation}
Again, flipping the `$>$' to '$<$' allows to decompose it as
\begin{align*}
    =&\frac{\pr{X_{\tau\!+\!k}<\Xp, X_\tau>\Xp, X_n<\Xp, X_0>\Xp, 1\leq n< \tau, 1\leq k <   \tau'}}{\pr{X_n<\Xp,X_0>\Xp, 1\leq n < \tau}-\pr{X_n<\Xp,X_0>\Xp, 1\leq n \leq \tau}}\\
    -&\frac{\pr{X_{\tau\!+\!k}<\Xp, X_\tau>\Xp, X_n<\Xp, X_0>\Xp, 1\leq n< \tau, 1\leq k\leq \tau'}}{\pr{X_n<\Xp,X_0>\Xp, 1\leq n < \tau}-\pr{X_n<\Xp,X_0>\Xp, 1\leq n \leq \tau}}\\
    =&\frac{\pr{X_{\tau\!+\!k}<\Xp, X_\tau>\Xp, X_n<\Xp,          1\leq n< \tau, 1\leq k <   \tau'}}{\copN[\tau-1]{}(p)-2\copN[\tau]{}(p)+\copN[\tau+1]{}(p)}\\
    -&\frac{\pr{X_{\tau\!+\!k}<\Xp, X_\tau>\Xp, X_n<\Xp,          0\leq n< \tau, 1\leq k <   \tau'}}{\copN[\tau-1]{}(p)-2\copN[\tau]{}(p)+\copN[\tau+1]{}(p)}\\
    -&\frac{\pr{X_{\tau\!+\!k}<\Xp, X_\tau>\Xp, X_n<\Xp,          1\leq n< \tau, 1\leq k\leq \tau'}}{\copN[\tau-1]{}(p)-2\copN[\tau]{}(p)+\copN[\tau+1]{}(p)}\\
    +&\frac{\pr{X_{\tau\!+\!k}<\Xp, X_\tau>\Xp, X_n<\Xp,          0\leq n< \tau, 1\leq k\leq \tau'}}{\copN[\tau-1]{}(p)-2\copN[\tau]{}(p)+\copN[\tau+1]{}(p)}\\
    =&\frac{\copN[\tau-1;\tau'-1]{}(p)-\copN[\tau  +\tau'-1]{}(p)}{\copN[\tau-1]{}(p)-2\copN[\tau]{}(p)+\copN[\tau+1]{}(p)} 
    - \frac{\copN[\tau  ;\tau'-1]{}(p)-\copN[\tau  +\tau'  ]{}(p)}{\copN[\tau-1]{}(p)-2\copN[\tau]{}(p)+\copN[\tau+1]{}(p)}\\
    -&\frac{\copN[\tau-1;\tau'  ]{}(p)-\copN[\tau  +\tau'  ]{}(p)}{\copN[\tau-1]{}(p)-2\copN[\tau]{}(p)+\copN[\tau+1]{}(p)} 
    + \frac{\copN[\tau  ;\tau'  ]{}(p)-\copN[\tau  +\tau'+1]{}(p)}{\copN[\tau-1]{}(p)-2\copN[\tau]{}(p)+\copN[\tau+1]{}(p)}\\
    =&\frac{\copN[\tau-1;\tau'-1]{}-\copN[\tau;\tau'-1]{}-\copN[\tau-1;\tau']{}+\copN[\tau;\tau']{}}{\copN[\tau-1]{}-2\copN[\tau]{}+\copN[\tau+1]{}}
    - \frac{\pi(\tau+\tau')}{\pi(\tau)},
\end{align*}
where 
\[
    \CopN[\tau;\tau']{p}=\mathcal{F}_{\libracket 0;\tau\!+\!\tau'\ribracket\backslash\{\tau\}}(F^{-1}(p),\ldots,F^{-1}(p)).
\]

\end{widetext}

\end{document}